\documentclass[a4paper]{jpconf}
\usepackage{graphicx}
\graphicspath{ {figures/} }

\begin{document}
\title{Shear viscosity at finite baryon densities}

\author{E McLaughlin$^1$, J Rose$^2$, T Dore$^3$, P Parotto$^4$, C Ratti$^5$ and J Noronha-Hostler$^3$}
\address{$^1$ Department of Physics, Columbia University, 538 W. 120th St., New York, NY 10027, USA}
\address{$^2$ Department of Physics,
Dr. Karl-Remeis Observatory  - Astronomical Institute,
Bamberg D-96049, Germany}
\address{$^3$ Illinois Center for Advanced Studies of the Universe, Department of Physics, University of Illinois at Urbana-Champaign, Urbana, IL 61801, USA}
\address{$^4$ University of Wuppertal, Department of Physics, Wuppertal D-42097, Germany}
\address{$^5$ Department of Physics, University of Houston, Houston, TX  77204, USA}
\ead{egm2153@columbia.edu}

\begin{abstract}
We use the excluded volume Hadron Resonance Gas (HRG) model with the most up-to-date hadron list to calculate $\eta T/w$ at low temperatures and at finite baryon densities $\rho_B$. This $\eta T/w$ is then matched to a QCD-based shear viscosity calculation of the QGP for different profiles of $\eta T/w$ across {T,$\mu_{B}$} including cross-over and critical point transitions. When compared to ideal hydrodynamic trajectories across {T,$\mu_{B}$}, we find that the $\eta T/w(T,\mu_B)$ profiles would require initial conditions at much larger baryon density to reach the same freeze-out point.
\end{abstract}

\section{Introduction}
Quark Gluon Plasma (QGP) has been observed in relativistic heavy ion collisions at both the Large Hadron Collider (LHC) and Relativistic Heavy Ion Collider (RHIC). In these relativistic heavy ion collisions at nearly zero baryon chemical potential, there is a smooth crossover phase transition from the QGP to HRG. However, experiments such as the RHIC Beam Energy Scan are looking for the existence of a critical point in the QGP to hadron gas phase transition at finite baryon densities. 

From comparisons of theory to experimental data we know the shear viscosity to entropy density ratio, $\eta /s$, of the QGP at zero baryon chemical potential is known to be very small, close to the KSS bound \cite{Kovtun:2004de}. 
It is not yet possible to calculate $\eta/s$ directly from lattice QCD, which necessitates the use of other approaches such as the HRG model \cite{NoronhaHostler:2008ju,NoronhaHostler:2012ug,Gorenstein:2007mw,Rischke:1991ke}, holography \cite{Kovtun:2004de} and transport models \cite{Rose:2017bjz}. 
However, in most of these models, one can only calculate $ \eta T/w(T,\mu_{B})$, the shear viscosity to enthalpy density ratio, in the hadronic phase (like in the cases of the HRG and transport models). 

Here we present a framework \cite{McLaughlin:2021dph} for connecting $\eta T/w(T,\mu_{B})$ in the HRG phase to a parametrized QCD-based description of $\eta T/w(T,\mu_{B})$ in the QGP phase. We construct four profiles of $\eta T/w(T,\mu_{B})$ across the QCD phase diagram: two cases of a fully crossover transition and two cases with a critical point (CP), the first at $(T,\mu_{B})=(143,350)$ MeV, matching the BEST collaboration EOS \cite{Parotto:2018pwx}, and the second at $(T,\mu_{B})=(89,724)$ MeV, to match the holographic prediction in \cite{Critelli:2017oub}. 

\section{Shear Viscosity in HRG Model}
We use the HRG model with particles from the 2016 Particle Data Group (PDG16+) list \cite{Alba:2017mqu}, where we take hadrons to be point like particles and can then calculate the total pressure of the system as:
\begin{equation}
    \frac{p(T,\mu_B,\mu_S,\mu_Q)}{T^{4}} = \sum\limits_{i}(-1)^{B_{i}+1} \frac{g_{i}}{2\pi^{2}}\int\limits_{0}^{\infty} p^{2}\ln{[1+(-1)^{B_{i}+1}e^{(-\frac{\sqrt{p^{2}+m_i^{2}}}{T}+ \tilde{\mu}_i)}]} dp
\end{equation}
To account for repulsive interactions in the hadron gas, we use the excluded volume approach, where we delegate a volume $v$ to each hadron and then recalculate thermodynamic quantities taking into account this excluded volume.
The shear viscosity of the HRG was then calculated at finite $\mu_{B}$
\begin{equation}\label{eqn:etqHRG}
    \eta^{HRG} = \frac{5}{64\sqrt{8}}\frac{1}{r^{2}}\frac{1}{n}\sum\limits_{i}n_{i}\frac{\int\limits_{0}^{\infty}p^{3}e^{-\sqrt{p^{2}+m_{i}^{2}}/T+\tilde{\mu}_i}dp}{\int\limits_{0}^{\infty}p^{2}e^{-\sqrt{p^{2}+m_{i}^{2}}/T+\tilde{\mu}_i}dp}
\end{equation}

\begin{figure}[ht]
    \centering
    \includegraphics[width=0.9\linewidth]{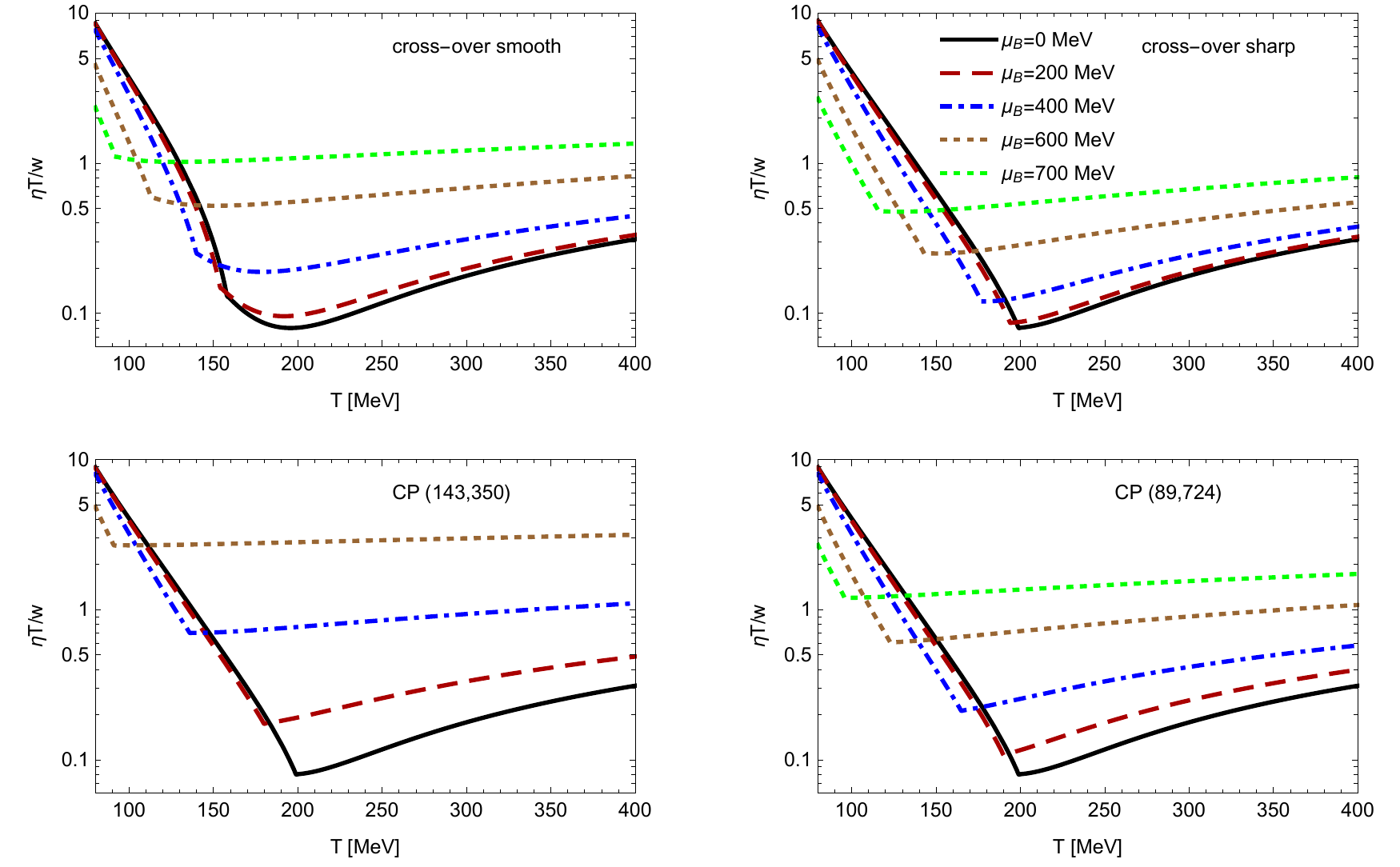}
    \caption{Temperature dependence of $\eta T/w(T,\mu_B)$ for all 4 transition profiles along slices of constant $\mu_B$}\label{fig:etaTw}
\end{figure}

The $ \eta T/w(T,\mu_{B})$ in the HRG phase is normalized to $ \eta /s=0.08$ at $\mu_{B}=0$ and splined to a parametrized $ \eta /s(T)$ from \cite{Dubla:2018czx} using a transition line across the $(T,\mu_{B})$ plane \cite{Borsanyi:2020fev}, the final results are shown in Fig.\ \ref{fig:etaTw}. 
In the smooth crossover transition we see a longer flat region in $\eta T/w$ that starts at around $\mu_{B} = 0$ and becomes more pronounced as you move to higher $\mu_{B}$. The matching of the QGP to HRG occurs at a lower temperature leading to a stronger $\mu_{B}$ dependence in the smooth crossover transition.
In comparison, in the sharp crossover the HRG to QGP matching occurs at a much higher temperature leading to $\eta T/w$ having a weaker $\mu_{B}$ dependence
So the shape and location in the {$T, \mu_{B}$} plane of the transition line from the deconfined to confined phase has a significant effect on how $\eta T/w$ is dependent on $\mu_{B}$.
For the  CP (143, 350) MeV transition there is also a strong $\mu_{B}$ dependence, since the critical point lies relatively close to $\mu_{B} = 0$, the transition line must fall steeply in temperature to pass through the critical point at (T, $\mu_{B}$) = (143, 350) MeV.


\section{Fluctuating Hydrodynamics}
\begin{figure}[ht]
  \begin{minipage}{0.5\linewidth}
    \centering
    \includegraphics[width=\linewidth]{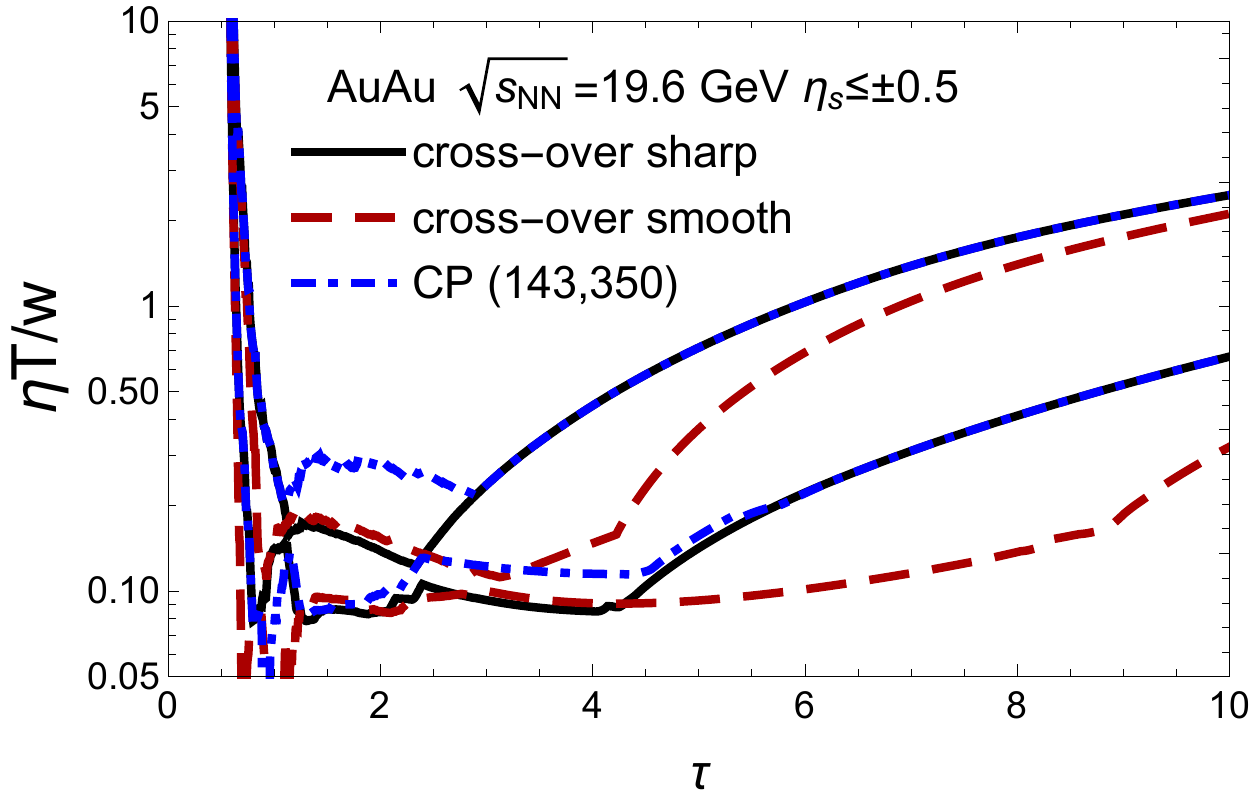}
  \end{minipage}
  \begin{minipage}{0.5\linewidth}
    \centering
    \includegraphics[width=\linewidth]{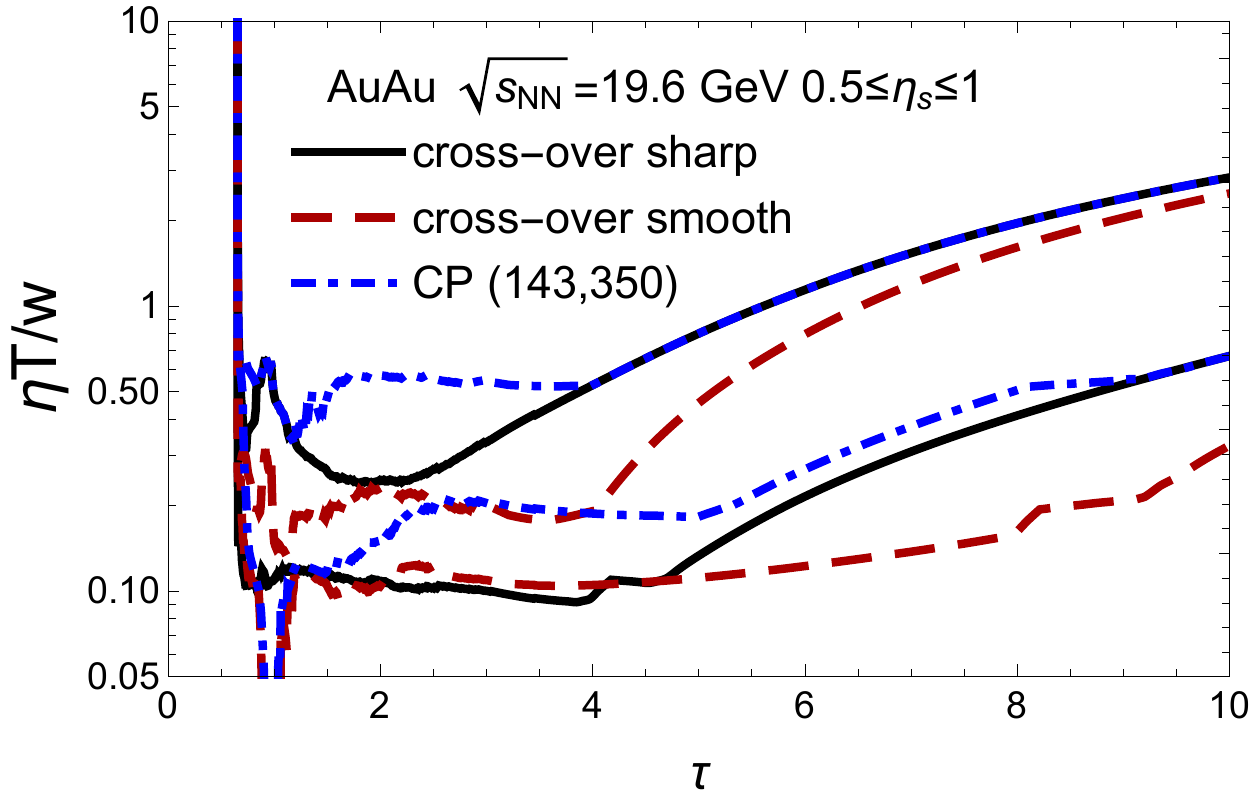}
  \end{minipage}
  \caption{Approximation of $\eta T/w$ at central (left) and forward (right) rapidity probed within relativistic viscous hydrodynamics vs. time (using the $\left(T,\mu_B\right)$ from a single hydrodynamic background from Ref.~\cite{Shen:2018pty})} \label{fig:etas_hydro}
\end{figure}

Variations in temperature and baryon chemical potential values probed by events at a specific beam energy can be caused by event-by-event fluctuations of the initial conditions, local temperature variations of the QGP, and entropy production due to the presence of viscosity. Different rapidity cuts may be used as a tool to probe these variations in T and $\mu_{B}$ throughout the collision space. 
Using the T and $\mu_{B}$ dependence of a single  hydrodynamic event modeling a Au-Au collision at 19.6 GeV from \cite{Shen:2018pty}, we construct the trajectory of $\eta T/w$ at central and forward space-time rapidity for three of our test cases, both crossover transitions and the (143, 350) MeV CP transition, as shown in Fig.\ \ref{fig:etas_hydro} where the mean temperature and baryon chemical potential is shown in solid lines, while $\eta T/w$ from T and $\mu_{B}$ values $\pm1$ standard deviation from the mean are shown in dashed lines.
In both the central and forward rapidity regions, the different profiles have similar minimum values for $\eta T/w$. This is most likely due to the fact that the average $\mu_{B}$ values in these regions are still relatively small ($\sim$200-300 MeV). At forward rapidity, $\eta T/w$ at early times is slightly higher and has more variation in the $\pm1$ standard deviation range, so we expect that this region is not at large $\mu_{B}$ but farther from equilibrium. 
\section{Influence on $T-\mu_{B}$ trajectories}
\begin{figure}[ht]
    \centering
    \includegraphics[width=0.5\linewidth]{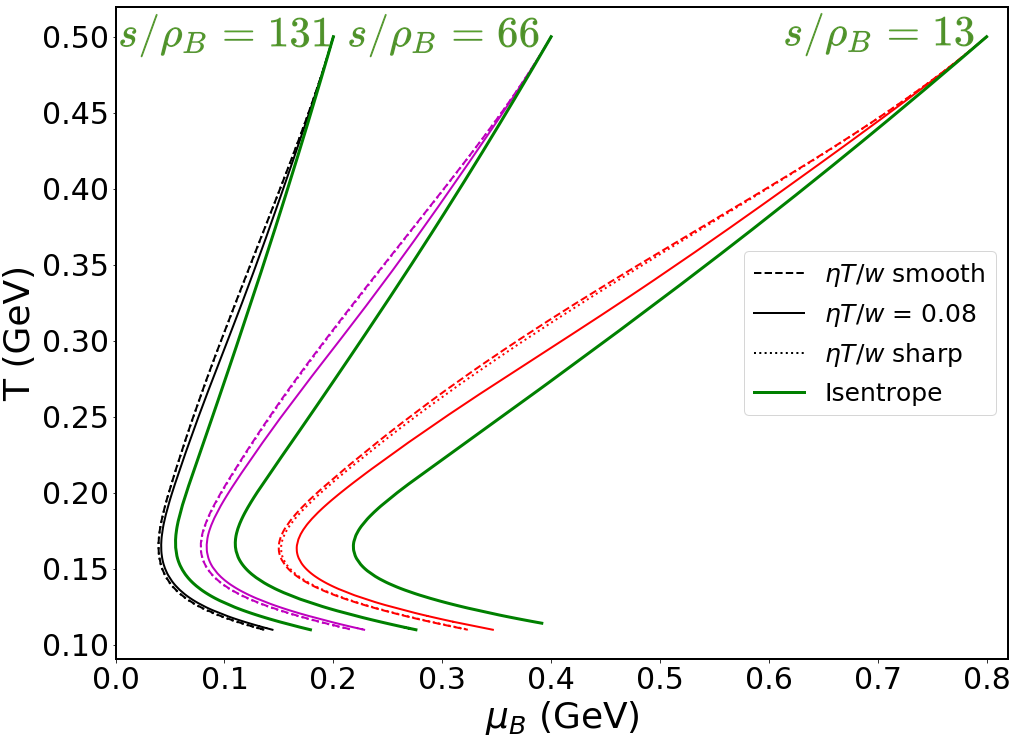}
    \caption{$\left\{T,\mu_B\right\}$ trajectories for various initial $s/\rho_B$ using $\eta T/w=0$, $\eta T/w=0.08$, or $\eta T/w$ profiles.}
    \label{fig:traj}
\end{figure}
We present the trajectories for hydrodynamic runs following the setup in \cite{Dore:2020jye}, beginning at T=500 MeV and various $\mu_{B}$ to compare the effect of using $\eta T/w(T,\mu_{B})$  to ideal hydrodynamics (isentropic trajectory) or a constant value for $\eta /s$.
Viscosity shifts to lower $\mu_B$ at freeze out because viscosity produces entropy as the system expands and cools. At large $\mu_B$ we see that the dependence of $\eta T/w$ on T and $\mu_{B}$ becomes more pronounced and the {$T,\mu_{B}$} trajectory deviates further from both ideal hydrodynamics and viscous hydrodynamics with a constant $\eta /s$, as shown in Fig. 3. In order to achieve the same freeze out conditions as the isentropic results, viscous hydrodynamic runs would need to have initial conditions that start at a higher $\mu_{B}$.

\section{Conclusions}
We have produced a framework for creating a functional $\eta T/w(T,\mu_{B})$ that can be used an input into relativistic viscous hydrodynamic codes simulating HIC at low RHIC energies where finite $\mu_{B}$ is of interest. In our framework, the profiles of $\eta T/w$ across (T, $\mu_{B}$) and minimum value of $\eta T/w$ along lines of finite $\mu_{B}$ are heavily determined by the transition line from QGP to HRG that one picks. The trajectories followed by a hydrodynamic system using our $\eta T/w(T,\mu_{B})$ profiles push the trajectories towards a lower $\mu_{B}$ at freeze out when comparing to both ideal hydrodynamics ($\eta /s = 0$) and a constant $\eta /s = 0.08$. The initial conditions of these viscous runs would need to be at higher $\mu_{B}$ to achieve the same freeze out conditions as ideal hydrodynamics. Furthermore, to better model HIC at low energies and finite baryon denisites, future steps should be taken to compute a description of $\eta T/w$ across the full (T,$\mu_{B},\mu_{S},\mu_{Q}$) space.

\section*{Acknowledgements}
This material is based upon work supported by the the US-DOE Nuclear Science Grant No. DE-SC0019175; the NSF under grants no. PHY-1654219 and PHY-1560077; and the DFG grant SFB/TR55.  

\section*{References}
\bibliography{all1}

\providecommand{\newblock}{}
\begin{thebibliography}{10}
\expandafter\ifx\csname url\endcsname\relax
  \def\url#1{{\tt #1}}\fi
\expandafter\ifx\csname urlprefix\endcsname\relax\def\urlprefix{URL }\fi
\providecommand{\eprint}[2][]{\url{#2}}

\bibitem{Kovtun:2004de}
Kovtun P, Son D~T and Starinets A~O 2005 {\em Phys. Rev. Lett.\/} {\bf 94}
  111601 (\textit{Preprint} \eprint{hep-th/0405231})

\bibitem{NoronhaHostler:2008ju}
Noronha-Hostler J, Noronha J and Greiner C 2009 {\em Phys. Rev. Lett.\/} {\bf
  103} 172302 (\textit{Preprint} \eprint{0811.1571})

\bibitem{NoronhaHostler:2012ug}
Noronha-Hostler J, Noronha J and Greiner C 2012 {\em Phys. Rev. C\/} {\bf 86}
  024913 (\textit{Preprint} \eprint{1206.5138})

\bibitem{Gorenstein:2007mw}
Gorenstein M~I, Hauer M and Moroz O~N 2007 {\em Phys. Rev. C\/} {\bf 77}
  214--220 (\textit{Preprint} \eprint{0708.0137})

\bibitem{Rischke:1991ke}
Rischke D~H, Gorenstein M~I, Stoecker H and Greiner W 1991 {\em Z. Phys. C\/}
  {\bf 51} 485--490

\bibitem{Rose:2017bjz}
Rose J~B, Torres-Rincon J~M, Sch\"afer A, Oliinychenko D~R and Petersen H 2018
  {\em Phys. Rev. C\/} {\bf 97} 055204 (\textit{Preprint} \eprint{1709.03826})

\bibitem{McLaughlin:2021dph}
McLaughlin E, Rose J, Dore T, Parotto P, Ratti C and Noronha-Hostler J 2021
  (\textit{Preprint} \eprint{2103.02090})

\bibitem{Parotto:2018pwx}
Parotto P, Bluhm M, Mroczek D, Nahrgang M, Noronha-Hostler J, Rajagopal K,
  Ratti C, Sch\"afer T and Stephanov M 2020 {\em Phys. Rev. C\/} {\bf 101}
  034901 (\textit{Preprint} \eprint{1805.05249})

\bibitem{Critelli:2017oub}
Critelli R, Noronha J, Noronha-Hostler J, Portillo I, Ratti C and Rougemont R
  2017 {\em Phys. Rev. D\/} {\bf 96} 096026 (\textit{Preprint}
  \eprint{1706.00455})

\bibitem{Alba:2017mqu}
Alba P {\em et~al.\/} 2017 {\em Phys. Rev. D\/} {\bf 96} 034517
  (\textit{Preprint} \eprint{1702.01113})

\bibitem{Dubla:2018czx}
Dubla A, Masciocchi S, Pawlowski J~M, Schenke B, Shen C and Stachel J 2018 {\em
  Nucl. Phys. A\/} {\bf 979} 251--264 (\textit{Preprint} \eprint{1805.02985})

\bibitem{Borsanyi:2020fev}
Borsanyi S, Fodor Z, Guenther J~N, Kara R, Katz S~D, Parotto P, Pasztor A,
  Ratti C and Szabo K~K 2020 {\em Phys. Rev. Lett.\/} {\bf 125} 052001
  (\textit{Preprint} \eprint{2002.02821})

\bibitem{Shen:2018pty}
Shen C and Schenke B 2019 {\em Nucl. Phys. A\/} {\bf 982} 411--414
  (\textit{Preprint} \eprint{1807.05141})

\bibitem{Dore:2020jye}
Dore T, Noronha-Hostler J and McLaughlin E 2020 {\em Phys. Rev. D\/} {\bf 102}
  074017 (\textit{Preprint} \eprint{2007.15083})

\end{thebibliography}

\end{document}